\documentstyle[prl,aps,preprint]{revtex}
\begin{document}
\draft
 \title{Long-Range Fluctuation-Induced Attraction of
       Vortices to the Surface in Layered Superconductors}
   \author{E.H.~Brandt,}
\address{Institut f\"{u}r Physik, Max-Planck-Institut f\"{u}r
Metallforschung, D-70506 Stuttgart, Germany}
   \author{R.G.~Mints, and I.B.~Snapiro}
\address{School of Physics and Astronomy, Raymond
and Beverly Sacler Faculty of Exact Sciences,\\ Tel Aviv
University, Tel Aviv 69978, Israel}
   \date{\today}
   \maketitle
   \begin{abstract}
It is shown that in extremely anisotropic layered superconductors
the interaction of vortex lines with a parallel planar surface,
which for straight lines along the $c$-axis decreases exponentially
over the in-plane penetration depth $\lambda$, becomes a long-range
dipole--dipole attraction when the vortex line is distorted randomly.
This novel long-range fluctuation-induced attraction
enhances the thermal fluctuations down to depths much larger than
$\lambda$ and may lead to flux creep towards the surface.
   \end{abstract}

Abrikosov vortex lines in layered superconductors have several
unusual properties as compared to vortex lines in isotropic or
weakly anisotropic superconductors. In particular, vortex lines
oriented perpendicular to the superconducting layers may be
considered as a stack of two-dimensional point vortices or pancakes
\cite{efet,arte,clem}. In the case of large anisotropy the pancakes
interact via a magnetic pair-potential, which parallel to the
layers decreases logarithmically and perpendicular to the layers
decreases exponentially. Most importantly, the interaction of
pancakes within
the {\it same} layer is {\it repulsive} while between different layers
it is {\it attractive} and reduced by a factor $s/2\lambda \ll 1$,
where $s$ is the layer spacing and $\lambda=\lambda_{ab}$ is the
penetration depth for the currents in the layers. As a consequence,
the interaction of two straight stacks of pancakes at distance
$\rho\gg\lambda$ decays as $\exp(-\rho/\lambda)$ since at long
distances the attraction and repulsion between the pancakes from both
stacks compensates almost exactly. Thus, the interaction of
two straight stacks is just the usual short-range repulsion of
Abrikosov vortices.
   \par
The repulsive and attractive interaction of point vortices
has a further important consequence, which to our knowledge has
not been pointed out previously, namely, the interaction of a
distorted pancake stack with a surface which is parallel to the stack.
Within the linear London theory the condition of zero perpendicular
current through a planar specimen
surface may be satisfied by adding the magnetic fields and currents
of image vortex lines \cite{bean,burl,kosh,mint}. Each vortex then
is attracted to its image since the images have opposite orientation
(antivortices). However, this short-range attraction applies only
when the vortex line is {\it perfectly straight}. As soon as the
vortex line is distorted, the compensation of repulsive and
attractive terms in the vortex-vortex interaction is no longer
ideal. As a consequence, randomly distorted vortex lines feel a
{\it long-range attraction to the surface}.
   \par
  In this paper we derive this novel long-range attraction and
discuss some of its consequences. We show that the result of the
isolated-layer model exactly coincides with the result of
anisotropic London theory in the limit $\lambda_c\gg\lambda$, where
$\lambda_c$ is the penetration depth for currents perpendicular to
the layers. The long-range attraction is, therefore, also present
in superconductors with finite anisotropy, where it is proportional
to $x^{-2}\exp(-2x/\lambda_c)$ for long distances $x$ between
vortex and surface. Since in high-$T_c$ superconductors
$\lambda_c$ typically is a macroscopic length, this surface
attraction is really of long range.
\par

We first give a simple physical interpretation of the
long-range fluctuation-induced attraction. Assume that only one of
the pancakes of a straight stack is displaced by a small distance $u$
away from the surface. This local distortion is formally described by
adding a pancake at the new position $x+u$ and an antipancake at the
equilibrium position $x$, which annihilates the original pancake.
The same procedure has to be done with the image stack situated at
the position $-x$ if the surface is at $x=0$ (see Fig.~\ref{fig1}).
The two pancake-antipancake pairs are dipoles with a strength
proportional to the displacement $u$ and a dipole--dipole interaction
energy proportional to $u^2/x^2$, where $1/x^2$ is the second
derivative of the pancake-pancake potential ($\ln x$). Therefore,
the distorted vortex line is attracted to its image, and thus to
the planar surface, by a long-range potential proportional to
$u^2/x^2$. This long-range interaction is in addition to the
short-range interaction of a perfectly straight vortex with its
image, which is proportional to
$K_0(2x/\lambda)\propto\exp (-2x/\lambda)$, where $K_0(x)$ is a
modified Bessel function.
   \par
   The interaction of a distorted vortex line with its image may be
calculated from the interaction of pancakes  separated
by ${\bf r}_{mn} = (x_m-x_n, y_m-y_n, z_m-z_n) =
 (x_{mn}, y_{mn}, z_{mn})$ \cite{feig,blat},
   \begin{equation}  
{\cal E} \approx \epsilon_0 \cases{
\displaystyle 2\ln( \rho_{mm} / \xi),    &$n=m$\cr
\displaystyle
 -{s\over 2\lambda} \exp\! \Big(\!- {|z_{mn}| \over \lambda}\Big)
 \ln\Bigl({\rho_{nm}\over\lambda}\Bigr), &$n\ne m$}
\label{e1}
   \end{equation}
with  $\epsilon_0 \! = s\Phi_0^2 /(4\pi \mu_0 \lambda^2)$,
$\rho_{mn}^2 \! = x_{mn}^2 + y_{mn}^2 $, and $z_m = ms$
where $s$ is the layer spacing and $\xi$ the core radius of the
pancake. For $z_m=z_n$ Eq.\ (1) applies to all distances
$\rho_{mn} > \xi$, but for $z_m\ne z_n$\,
$\rho_{mn} \gg \lambda$ was assumed.  The total energy of a
vortex line is composed of its self-energy and the interaction
with its image line of opposite orientation, namely,
   \begin{eqnarray}  
 {1\over 2} \sum_m \sum_n [\,
     {\cal E}(x_m-x_n,\, y_m-y_n,\, z_m-z_n) \nonumber \\
   - {\cal E}(x_m+x_n,\, y_m-y_n,\, z_m-z_n)\,]\,,
   \end{eqnarray}
where the sums are over the pancakes of the {\it real} vortex in
the superconducting half space $x>0$, thus $x_m, x_n >0$.
For a distorted vortex line parallel to the surface $x=0$ we
define pancake displacements ${\bf u}_m = {\bf u}_m(z_m) =
(u_{xm}, u_{ym})$  by writing  $x_m =x +u_{xm}$ and $y_m =u_{ym}$.
Keeping only the terms quadratic in the ${\bf u}_m$ we obtain
from (2) the linear elastic energy of the vortex.

   We first consider random and isotropic displacements with
ensemble averages
 $\langle u_{xm} \rangle = \langle u_{ym} \rangle = 0$,
         $ \langle u_{xm}u_{xn} \rangle =
          \langle u_{ym}u_{yn} \rangle = f(|m-n|) $,
         $\langle u_{xm}u_{yn} \rangle = 0$. The long-range
interaction with the image line for $2x \gg \lambda$ then becomes
from (1) and (2)
   \begin{equation}  
   E_{\rm int}= -{\Phi_0^2\, L\, s\over 64\pi \mu_0 \lambda^3 x^2}
   \sum_l \exp\!\Big( \! -{ s \over \lambda} |l| \Big)\,
   \langle ({\bf u}_l- {\bf u}_0 )^2 \rangle \,,
  \end{equation}
where $L$ is the vortex length. This fluctuation-induced
interaction is attractive  and depends on the relative
displacements $\langle ({\bf u}_l- {\bf u}_0 )^2 \rangle$ over
a vortex length of order $\lambda$. Like a dipole--dipole
interaction it decreases as $1/x^2$. Since $s \ll \lambda$, one
may approximate the sum in (3) by   
   \begin{equation}  
   E_{\rm int} = - {\Phi_0^2 L \over 32 \pi \mu_0 \lambda^3 x^2 }
   \int_0^\infty \!\! dz \,
   \exp\!\Big( \! -{ z \over \lambda} \Big)\, g(z) ,
   \end{equation}
where $g(z)$ is the correlation function
   \begin{equation}  
   g(z)  = \langle [ {\bf u}(z)- {\bf u}(0) ]^2 \rangle \,.
   \end{equation}
The integral (4) shows that the long-range interaction does not
depend on the layer separation $s$.

   For a more comprehensive calculation we use the general expression
for the two-pancake interaction \cite{feig,blat},
   \begin{equation}  
   {\cal E} = {\Phi_0^2 s^2 \over \mu_0} \int\! {d^3 k \over 8\pi^3}
   { 1 \over 1+k^2\lambda^2}\, { k^2 \over q^2}\,
   e^{\, i\displaystyle {\bf kr}_{mn}}
   \end{equation}
where ${\bf k} = (k_x, k_y, k_z)$, ${\bf q}=(k_x, k_y)$, $p=k_z$,
thus $q^2 = k_x^2 +k_y^2$ and  $k^2 = q^2 +p^2$, and as above
${\bf r}_{mn} = (x_{mn}, y_{mn}, z_{mn})$. Since (6) is valid
for all ${\bf r}_{mn}$, we may obtain the total elastic energy of
the distorted vortex line $E_{\rm tot} = E_{\rm self}+E_{\rm int}$
from (2) and (6). Expanding this to quadratic terms in ${\bf u}$,
introducing the Fourier transform
   \begin{equation}  
   {\bf u}(z_l) = \int\! {dp \over 2\pi}\,\, {\bf \tilde u}(p)\,\,
   e^{\displaystyle ipz_l}
   \end{equation}
and using  $s\sum \exp(ipz_l) = 2\pi \delta(p)$
(for $|p| \le \pi/s$) and $\int(dp/2\pi)\, |{\bf \tilde u}(p)|^2 =
  \int \! dz\, {\bf u}(z)^2 = \langle {\bf u}^2 \rangle L $
we obtain
   \begin{eqnarray}  
   E_{\rm tot} &=& {\Phi_0^2 \over 4\mu_0} \int\! {d^3 k \over 8\pi^3}
   \, |{\bf \tilde u}(p)|^2 \, (f_{\rm self} + f_{\rm int}) \,, \\
   f_{\rm self} &=& {k^2 \over 1+k^2\lambda^2}
                - {q^2 \over 1+q^2\lambda^2} \,,   \\
   f_{\rm int}  &=& \Big( {k^2 \over 1+k^2\lambda^2}\, {k_x^2
     \over q^2} - {1 \over 2}f_{\rm self} \Big)\,
     e^{\displaystyle 2ik_x x}.
   \end{eqnarray}

   Remarkably, exactly the same result (8)-(10) is obtained from the
anisotropic London theory in the limit $\lambda_c \to \infty$.
Namely, the interaction energy of two London vortices at
positions ${\bf r}_1(z)$ and ${\bf r}_2(z)$ is
   \begin{equation} 
   \int\!\! d{\bf r}_{1\alpha} \int\!\! d{\bf r}_{2\beta} \,\,
   V_{\alpha\beta} ({\bf r}_1 -{\bf r}_2) \,,
   \end{equation}
where the anisotropic interaction
$V_{\alpha\beta} ({\bf r}_1 -{\bf r}_2)$ ($\alpha,\beta = x,y,z$)
is given in Ref. \cite{bra1} as a Fourier integral over
$V_{\alpha\beta}({\bf k})$. In the limit $\lambda_c \to \infty$,
the general expression $V_{\alpha\beta}({\bf k}) =
 (\Phi_0^2 /\mu_0)(1+k^2\lambda^2)^{-1} g_{\alpha\beta}({\bf k})$
simplifies to a diagonal matrix with $g_{xx}=k_x^2/q^2$,
$g_{yy} = k_y^2/q^2$, and $g_{zz} =1$ if $z$ is along the
$c$-axis of the uniaxial superconductor. Inserting this
 $V_{\alpha\beta}({\bf k})$  into (11) and integrating over the
vortex and its image, we reproduce the result (8)-(10) of the
pancake approach.

   From Eq.\ (10) one reproduces the long-distance
interaction (3), (4). Moreover, from the self-interaction (9) the
line tension $P$ of an isolated flux line \cite{bra2} or stack of
pancakes is obtained,
   \begin{equation} 
  P(p) = { \Phi_0^2 \over 8\pi\,\mu_0\, \lambda^2} \,
         { \ln(1+p^2\lambda^2) \over p^2\, \lambda^2 } \,,
   \end{equation}
which determines the linear elastic self-energy of a distorted
vortex line,
   \begin{equation} 
   E_{\rm self} = {1\over 2} \int\! {dp \over 2\pi}\,
                   p^2 P(p)\, |{\bf \tilde u}(p)|^2 .
   \end{equation}
In real space this energy looks similar to Eq.\ (3),
   \begin{equation}  
   E_{\rm self} = {\Phi_0^2 L \over 16 \pi \mu_0 \lambda^4 }
   \sum_{l\ne 0} \exp\! \Big( \! -{ s \over \lambda} |l| \Big)\,
   {\langle ({\bf u}_l- {\bf u}_0 )^2 \rangle \over |l| } ,
   \end{equation}
 and since $s \ll \lambda$ it can be approximated as
   \begin{equation} 
   E_{\rm self} = {\Phi_0^2 L \over 8 \pi \mu_0 \lambda^4 }
   \int_0^\infty \!\! dz \,
   \exp\!\Big( \! -{ z \over \lambda} \Big)\, {g(z) \over z} .
   \end{equation}
The similarity of (4) and (15) leads to the following useful
relationship. If the correlation function (5) increases
algebraically, $g(z) = {\rm const} \cdot |z|^\gamma$, one has
(still for $2x \gg \lambda$),
   \begin{equation}  
     E_{\rm int} = - (\gamma \lambda^2 /4 x^2)\, E_{\rm self} \,.
   \end{equation}
For example, a flux line diffusing in a random pinning potential
may exhibit $\gamma = 1$, thus $E_{\rm int} =  -(\lambda^2/4x^2)
 E_{\rm self}$ is not a very small correction to $E_{\rm self}$.

   To obtain the thermal fluctuations one has to consider the
pronounced dispersion  (``non-locality'') of the line tension
$P \propto 1/p^2$ (12) at large $p \gg \lambda^{-1}$. This means
that a single displaced vortex ``feels'' a parabolic potential
of curvature $p^2 P(p) \approx \Phi_0^2 \ln(\alpha\lambda/ s)/
 (4\pi\mu_0 \lambda^4) $ (with $\alpha =1.16$ for $s\ll \lambda$)
originating from the global interaction with all pancakes within
a distance of order of $\lambda$ along the stack. Therefore, the
self-energy (3) caused by short-wavelength fluctuations is simply
   \begin{equation} 
   E_{\rm self} = {\Phi_0^2 L\, \ln(\alpha \lambda/s) \over
    8 \pi \mu_0 \lambda^4 }\, \langle {\bf u}^2 \rangle \,.
   \end{equation}
Each pancake only
weakly feels the usual ``local'' (non-dispersive) line tension
that originates from the nearest-neighbor interaction, which
here is strongly reduced due to the absence of Josephson coupling
between the layers. For long tilt wavelengths $2\pi/q \gg \lambda$,
however, the line tension (12) of the pancake stack is local,
$P = \Phi_0^2 /(8\pi \mu_0 \lambda^2)$, as can be seen also by
inserting a uniform tilt ${\bf u}_l -{\bf u}_0 \propto l$ in (14).

  In terms of the Fourier transform (7) the correlation function
g(z) in (4) and (15) may be expressed as
   \begin{equation}  
   g(z) = {2 \over L} \int \! {dp \over 2\pi} \,
   \langle |{\bf \tilde u}(p)|^2\rangle\, [\,1- \cos(pz) \,] \,.
   \end{equation}
Writing the self energy (11) for an isolated vortex of finite
length $L$ as a sum over discrete modes with wave vectors $p_n$
and amplitudes ${\bf u}_n$ and ascribing to each mode the average
thermal energy $k_B T$ (since ${\bf u}_n$ has two components),
one obtains $\langle| {\bf \tilde u}(p) |^2\rangle$ and with (18),
   \begin{equation}  
 g(z) = k_B T\, {32 \pi \mu_0 \lambda^4 \over \Phi_0^2 }\! \int \!
   {dp \over 2\pi } \, {1 -\cos(pz) \over \ln(1+ p^2 \lambda^2) } .
   \end{equation}

The function $g(z)$ is practically constant in the interval relevant
in Eqs.~(4) and (15). One has $g(0)=0$ and for $|z| \gg s$ one finds
$g(z) \approx c_1 + c_2 |z| /\lambda$ with $c_2 \ll c_1$. For the
example $\lambda/s=100$ some values of the integral in (19),
times $\lambda$, are
 0, 4.71, 5.07, 5.16, 5.25, 5.29, 5.42, 5.53, 5.68, 5.84, 6.10, 6.85
for $z/s=0$, 1, 2, 3, 4, 5, 10, 20, 50, 100, 200, 500.
The constancy of $g$ at $z$ up to a few $\lambda$ means that
the thermal fluctuations  of the pancakes of
one vortex line are nearly {\it uncorrelated} at small
wavelengths \cite{clem,bula}, and thus the cutoff length of
several $s$, or the effective cutoff $|q| \le \alpha/s$,  required
to obtain expressions like Eq.\ (17) {\it is not crucial}. This
``elastic independence'' of the pancakes originates from the strong
dispersion of the line tension $P(p)$ (12).
For $|z| \gg 2\pi\lambda$ and for a usual string with
non-dispersive $P$ the thermal random walk yields $g(z)\propto|z|$.

The expression for g(z) (19) was derived from the line energy
$E_{\rm self}$ (13), which does not depend on the vortex distance
$x$ from the surface. The correct calculation, however, has to
consider the total energy $E_{\rm self} + E_{\rm int}$. Since
the interaction with the surface $E_{\rm int}$ depends on  $x$,
and the equipartion theorem requires constant energy $k_B T$ per mode,
we find that the fluctuation amplitude depends on $x$.
Explicitly we obtain from (4) and (15) for a vortex line at
$x \gg \lambda/2$ the short-wavelength thermal fluctuations,
{\it i.e.}, the value $g(z)$ at $s < |z| \alt \lambda$,
   \begin{equation}  
   \langle {\bf u}^2 \rangle \approx {2 \, \lambda^2
     \over \ln(\alpha\lambda/s)}\, {k_BT \over \epsilon_0}\,
   \Big[ 1 + {\lambda^2 \over 4 x^2 }\,
   {1 \over \ln(\alpha \lambda/s) }\, \Big] \,.
   \end{equation}
 So we have the interesting result that
the correction to the thermal fluctuations decreases away from
the surface {\it only as a power law}. This increase of
 $\langle {\bf u}^2 \rangle$, which in the total energy exactly
compensates the spatial dependence of the term $E_{\rm int}$,
originates from the softening of the flux-line lattice
near the surface.

   Considering $u_x$ and $u_y$ separately, we find that the
surface-caused softening of a vortex line is indeed {\it isotropic}
at large $x$ but becomes {\it anisotropic} for $x < \lambda$.
The general expression obtained from (2) and (6)  shows that
to a good approximation the contributions of each displaced
pancake to the elastic energy add {\it independently}. Each
pancake contribution is composed of the interaction between six
objects, namely, the displaced pancake
and the hole (antipancake) it leaves in the vortex line, their
images in the {\it same} layer, the {\it undisplaced straight}
vortex line, and its image. Explicitely one has
   \begin {eqnarray} 
  E_{\rm tot} &=& {\Phi_0^2\, L \over 8\pi\mu_0 \lambda^4} \Big[\,
  a(x) \langle u_x^2\rangle + b(x) \langle u_y^2\rangle \,\Big],
                                                     \nonumber\\
  a(x) &=& \ln{\alpha\lambda \over s} - {\lambda^2 \over 4x^2}
  - K_0''\Big({2x \over\lambda}\Big)\,,              \nonumber\\
  b(x) &=& \ln{\alpha\lambda \over s} - {\lambda^2 \over 4x^2}
  - {\lambda \over 2x} K_0'\Big({2x \over\lambda}\Big) \,,
   \end{eqnarray}
with $\alpha =1.16$. For large $x \gg \lambda/2$ one has $a=b=
 \ln(\alpha\lambda/s)  -\lambda^2 /4x^2$, and for $x\ll \lambda/2$,
 $a=b - \lambda^2 / 2x^2$, $b=\ln(\alpha\lambda/s) $,
 since the derivatives are $K_0'(r) = -1/r$ and $K_0''(r) = 1/r^2$
 for $r\ll 1$. The thermal fluctuations
 \begin{equation} 
 \langle u_x^2 \rangle = {2k_BT \lambda^2 \over\epsilon_0\, a(x)},~~~
 \langle u_y^2 \rangle = {2k_BT \lambda^2 \over\epsilon_0\, b(x)}
 \end{equation}
become thus {\it asymetric} near the surface, with
$\langle u_x^2 \rangle / \langle u_y^2 \rangle = b/a > 1$. Note that
the curvature $a$ in (21) turns negative at small $x$. To check the
stability of a pancake stack close to the surface and to obtain the
correct pictures of penetration and exit, one has to
add the (always positive) stray-field energy \cite{bra3,marc}
and the restoring force of the current density $j(x)$ that
originates from the Meissner surface current and from other vortices,
 replacing  $a$ by $a + (4\pi\mu_0\lambda^4 / \Phi_0) j\,'(x)$.
Work in this direction is under way.

   Large fluctuations and local tilt of flux lines can
be caused by pinning. Random pinning {\it forces} on a single
flux line would cause square displacements
diverging proportional to the number of forces. More realistic
random pinning {\it potentials} lead to finite  vortex
displacements \cite{blat,giam}.
For a crude estimate assume that random pins of density $n_p$ are
so strong that the vortex line wanders an average
distance squared of order $(4n_p s)^{-1}$ as it passes to the
next layer. This yields $g(z) \approx z/(n_p s^2)$, and
the interaction (4) with the surface becomes
   \begin{equation}  
   E_{\rm int} = -{\Phi_0^2\, L \over 32 \pi \mu_0 \lambda^3
   n_p s^2} \, { \lambda^2 \over x^2 } =
    - {\lambda^2 \over 4x^2 }\, E_{\rm self}\,.
   \end{equation}
 The resulting  long-range force
$-dE_{\rm int}/dx$ in principle may drive the vortex line to the
surface. It has to be added to geometric forces which are exerted
on the vortex line by surface screening currents when the
specimen has a non-ellipsoidal cross section and is not infinitely
thick \cite{schu,zeld}.

   The elastic energy (8)-(10) of a {\it single} flux line may
be generalized to the elastic energy of the distorted
flux-line {\it lattice} in the half space $x>0$ by replacing in it
$| {\bf \tilde u}(q) |^2$ by the three-dimensional Fourier transform
$| {\bf \tilde u(k)} |^2$ of the lattice displacements \cite{sudb}.
The complete expression for
$E_{\rm tot}$ slightly differs  from our approximation
(8), which applies if  $\langle u_x^2 \rangle =
 \langle u_y^2 \rangle = {1\over 2} \langle {\bf u}^2 \rangle $
 and $\langle u_x u_y \rangle =0$.
The presence of a flux-line lattice of density $B/\Phi_0$
increases the fluctuation-induced attraction to the surface (3)
when the displacements in the same layer are correlated
over a radius $\rho_{\rm corr} > (\Phi_0/B)^{1/2}$. The
interaction per vortex will then increase by a factor of order
of $\pi \rho_{\rm corr}^2 B/\Phi_0$.

Similar expressions for the elastic energy of\, isotropic and
anisotropic vortex lattices near a planar surface are derived
in Refs. \cite{bra3,marc}. In isotropic superconductors a
long-range attraction of the type (3) is absent, and in anisotropic
superconductors the attraction is reduced by a factor of
$\exp(-x/\lambda_c)$. Formally, the long range is due to the
non-cutoff factor  $1/q^2$ in Eq.\ (10).

  In conclusion, we have shown that in extremely anisotropic
layered superconductors there is a long-range interaction between
the sample surface and a distorted vortex line parallel to the
surface and to the $c$-axis.
This interaction causes a spatial variation of the thermal
fluctuations even at distances much larger than the in-plane
London penetration depth, which might affect the melting
process of the vortex lattice \cite{nels} and the evaporation
of vortex lines into independent pancake vortices \cite{clem,bula}.
 In the presence of sufficiently strong random pinning, a
fluctuation-induced long-range force attracts the distorted vortex
line to the surface. This additional force means a bulk current
density far inside the superconductor and may lead to flux creep
towards the surface.

  We acknowledge support from the German-Israeli Foundation for
Research and Development, Grant \mbox{\# 1-300-101.07/93.}

\begin{figure}
\caption{{\it Left:\/} A distorted vortex line and its image
composed of pancakes ($\uparrow$) and antipancakes ($\downarrow$).
{\it Right:\/}  The two dipoles generated by the displacement
cause a long-range attraction between the distorted vortex and
the surface  (indicated by a vertical bold line). }
\label{fig1}
\end{figure}

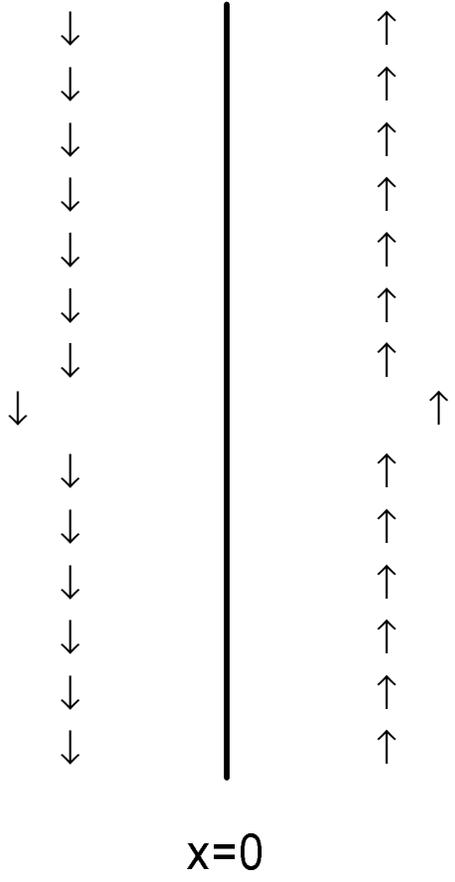
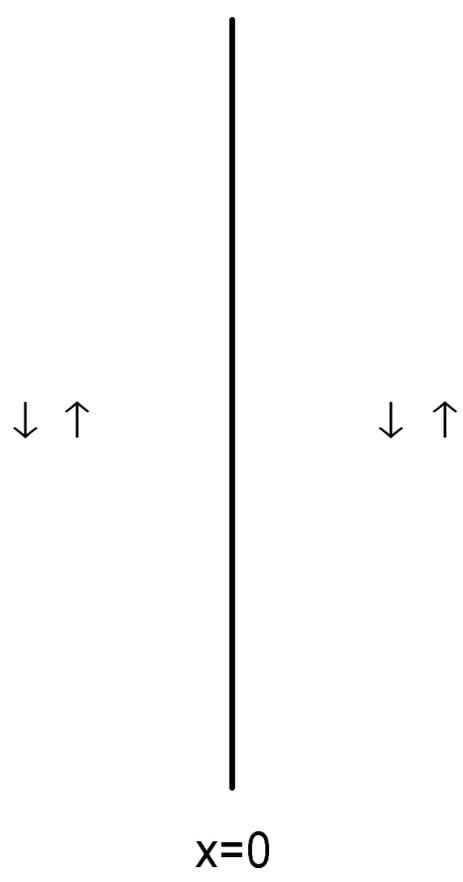
\end{document}